\documentclass[usenatbib]{iau_FM}
\usepackage{graphicx}
\usepackage{natbib}
\usepackage{hyperref}

\DeclareRobustCommand{\ION}[2]{%
\relax\ifmmode
\ifx\testbx\f@series
{\mathbf{#1\,\mathsc{#2}}}\else
{\mathrm{#1\,\mathsc{#2}}}\fi
\else\textup{#1\,{\mdseries\textsc{#2}}}%
\fi}

\newcommand{\Ssfr}{$\Sigma_{\rm SFR}$\,}
\newcommand{\Sst}{$\Sigma_{\rm *}$\,}
\newcommand{\Sgas}{$\Sigma_{\rm mol}$\,}

\title[The relations that regulates the SFR] 
{The local and global relations between
$\Sigma_\star$ , $\Sigma_{\rm SFR}$ and $\Sigma_{\rm mol}$ that regulate star-formation}

\author[Sebastián F. Sánchez]   
{Sebastián F. Sánchez$^1$
 \and Daysi C. Gómez Medina$^2$
 \and J.K. Barrera-Ballesteros$^1$
 \and L. Galbany$^3$
 \and A. Bolatto$^4$
 \and T. Wong$^5$
 }

\affiliation{$^1$Instituto de Astronomía,\\ Universidad Nacional Autónoma de México, \\A. P. 70-264, C.P. 04510, Ciudad de México, Mexico
\\ email: {\tt sfsanchez@astro.unam.mx} \\[\affilskip]
$^2$Facultad de Ciencias Espaciales, \\ 
Universidad Nacional Autónoma de Honduras\\[\affilskip]
$^3$ Institute of Space Sciences (ICE, CSIC), Campus UAB, Carrer de Can Magrans, s/n, E-08193 Barcelona, Spain.\\[\affilskip]
$^4$Department of Astronomy, University of Maryland, \\ College Park, MD 20742, USA\\[\affilskip]
$^5$Department of Astronomy, University of Illinois,\\ Urbana, IL 61801, USA}

\pubyear{2015}
\jname{Astronomy in Focus, Volume 1} 
\editors{Piero Benvenuti, ed.}
\begin{document}

\maketitle

\begin{abstract}


Star-formation is one of the main processes that shape galaxies, defining its stellar population and metallicity production and enrichment. It is nowadays known that this process is ruled by a set of relations that connect three parameters: the molecular gas mass, the stellar mass and the star-formation rate itself. These relations are fulfilled at a wide range of scales in galaxies, from galaxy wide to kpc-scales. At which scales they are broken, and how universal they are (i.e., if they change at different scales or for different galaxy types) it is still an open question. We explore here how those relations compare at different scales using as proxy the new analysis done using Integral Field Spectroscopy data and CO observations data from the EDGE-CALIFA survey and the AMUSSING++ compilation.
\end{abstract}

\firstsection 
\section{Introduction}

One of the main processes that shape a galaxy is indeed the
star-formation, defining its very existence. A certain set of physical conditions are required to trigger the thermonuclear
reactions that define a star, starting from the fueling of atomic gas, its cooling and condensation to generate molecular gas clouds, and the fragmentation of those clouds and its collapse to reach the required physical conditions to ignite them.
All those conditions shape a set of relations between the star-formation rage itself, the gas mass and the stellar mass content. Those relation are nowadays evident that regulates the star-formation process.

\citet{schmidt59} first proposed a relation between the SFR and the density of gas
in a certain volume, based on theoretical considerations. This relation was later
analyzed in \citet{schmidt68}, but it was not until \citet{kennicutt1998} that it
was not expressed in its current form: a log-log (or power law) relation between
(\Ssfr and \Sgas), i.e., two intensive parameters that do not depend on the size of galaxies. In \citet{kennicutt98} it was predicted an slope of $\sim$1.4 for this relation, based on free-fall time considerations for a self-collapsion cloud.
Although it was first derived as a relation between global intensive parameters, it
is now known to hold above the typical scale of large molecular clouds ($\sim$500 pc),
i.e., described as a relation for spatially resolved sub-galactic structures\citep[e.g., the
rSK-law][]{wong02,Kennicutt07}. Both the global and the resolved relations present a similar dispersion, of the order of$\sim$0.2 dex \citep{bigiel08,leroy13}.  However, 
there is discrepancy in the slope with respect to the theoretical considerations, being nowadays more near to a slope $\sim$1 \citep[e.g.][]{ARAA,sanchez20}.

The relation between the star-formation rate (SFR) and the stellar mass (M$_*$) has been explored uncovered more recently, based on the exploration of large galaxy surveys such as the SDSS \citep[][]{york+2000}. This relation, known as the Star-Formation Main Sequence \citep[e.g.][]{brin04,renzini15} is a tight correlation ($\sigma \sim$0.25 dex) between the logarithm of both quantities, with a slope near to one, observed only for star-forming galaxies (i.e., those which ionization is clearly dominated by the effect of young massive OB stars). It has been observed in a wide range of redshifts, explored in more detail in the nearby  Universe ($z\sim0$), but present up to redshift 1-2 \citep[e.g.][]{speagle14,rodriguez16}. It presents a clear evolution, at least in its zero-point, as galaxies increase their stellar masses and decreases their SFR from early cosmological times to the present epoch. Two almost simultaneous studies, \citet{sanchez13} and \citet{wuyts13}, proposed the existence of resolved version of this relation that holds down to $\sim$1kpc scales, the rSFMS, observed as a tight relation between the logarithms of the \Ssfr and \Sst. Detailed explorations, like the one presented by
\citet{cano16}, using data from the CALIFA survey \citep{califa}, 
has estabilished definetively its shape and characteristics. A possible dependence of that relation with other properties of the galaxies, such as the morphology, has been explored by different authors \citep[e.g.][]{rosa16,catalan17,mariana19,jairo19}.

A third relation between the molecular gas and stellar masses has been described for SFGs too \citep[e.g.][]{saint16,calette18}. Like the previous two it is a tight relation with a slope near to one. This relation, known as the Molecular Gas Main Sequence (MGMS), has been less frequently explored in the literature. It has not been until recent times that its resolved counterpart has been explored at a kpc scale, \citep[rMGMS][]{lin19}, as a relation between the corresponding intensive parameters: \Sgas
vs. \Sst, using a combination of IFS data provided by the MaNGA survey \citep{bundy15}, and CO-mapping provided by the ALMAQUEST compilatoin \citep[][]{lin20}.

The connection between the global (galaxy-wide) and resolved (kpc-scale) relations, its universality, the possible dependence with additional parameters, and
the interconnection or hierarchy between the three relations is a topic of study
that it is still open. \citet{bolatto17}, first shown that a simple parametrization
of the global intensive SK-relation follows the same distribution observed in the
rSK between \Ssfr and \Sgas. More detailed explorations have shown the same
correspondence between the global extensive SFMS and the local/resolved rSFMS \citep{pan18,mariana19}. Finally, \citet{sanchez20} and \citet{sanchez2021b}, have shown that the global intensive relations (i.e., when \Ssfr, \Sgas and \Sst are measured as average quantities galaxy wide) and the local/resolved relations (i.e., when those parameters are measured in star-forming regions at a kpc-scale), present the same distributions, the same slopes and similar zero-points. These explorations were possible due to the unique combination of spatially resolved observations of large sample of galaxies using both optical IFS and CO millimetric data, and the use of recent tracers of the gas content based on dust attenuation values \citep[e.g.][]{jkbb20}.

In this manuscript we present new explorations of the connection between the global
and local relations using recent results obtained using IFS data on an unique compilation of MUSE \citep{bacon01} data for a sample of galaxies in the nearby Universe. We use these results to discuss on the nature of these relations and the physical reason for their existence. Throughout this article we assume the standard $\Lambda$ Cold  Dark Matter cosmology with the parameters: $H_0$=71 km/s/Mpc, $\Omega_M$=0.27, $\Omega_\Lambda$=0.73.

\begin{figure}
    \centering
    \includegraphics[width=\textwidth]{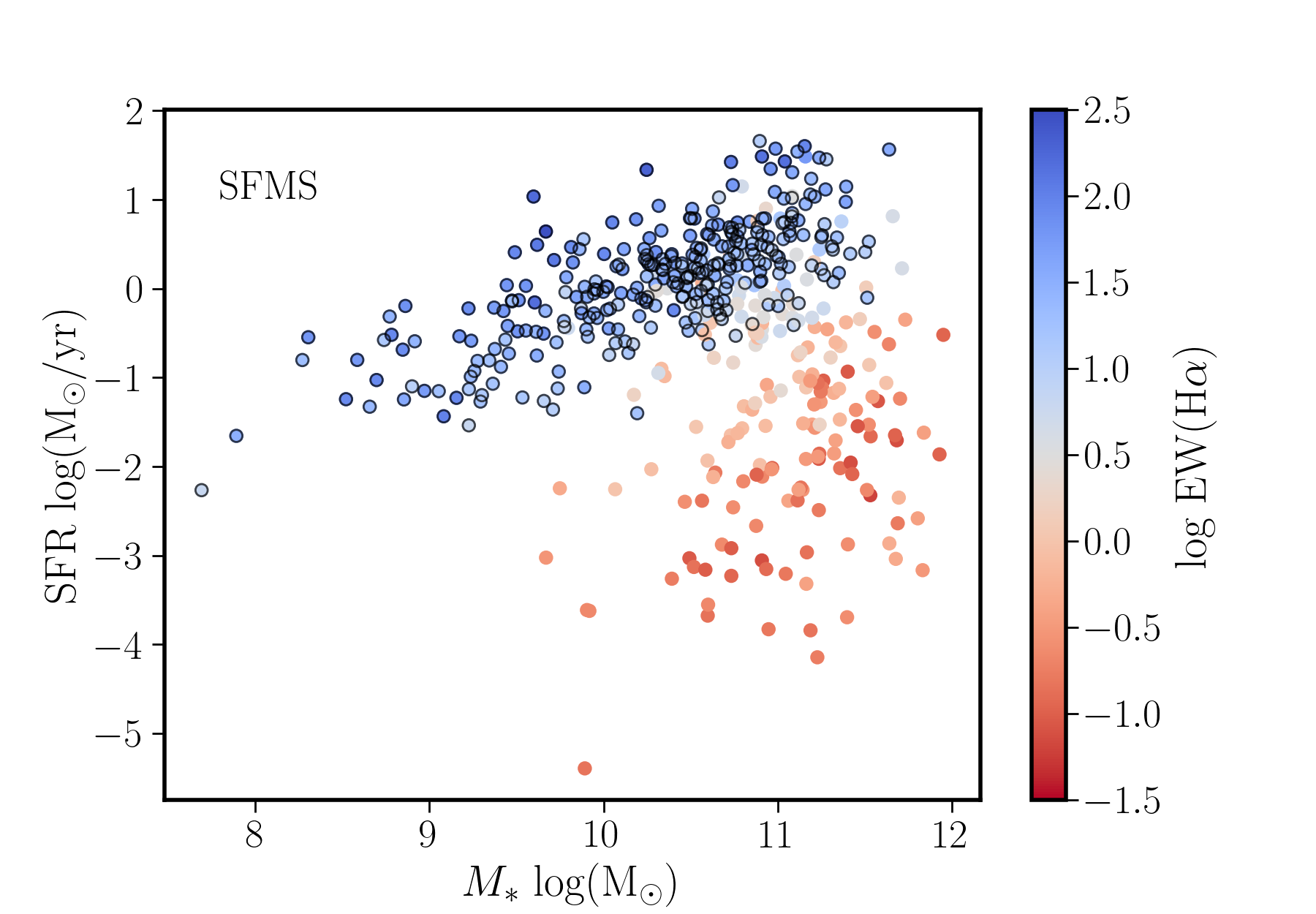}
    \caption{Distribution of the SFR along M$_*$ for AMUSING++ galaxy compilation. Each solid-circle corresponds to one particular galaxy, color-coded by the EW(H$\alpha$) at the effective radius. Black open-circles indicate that SFGs selected from the original sample.}
    \label{fig:sample}
\end{figure}

\section{Sample and Data}

We made use of the data provided by AMUSING++ compilation \citep{carlos20}, a set of observations using MUSE on galaxies in the nearby Universe ($z\sim$0.03), that comprises $\sim$600 galaxies (in the current compilation). These observations comprises the galaxies included in the AMUSING survey (PI: J. Anderson), and data retrieved from the ESO archive. All galaxies were selected to match their optical extension with the FoV of the MUSE instrument ($\sim$1 arcmin$^2$), in a way that it covers between 1.0 and 2.0 effective radius. We exclude edge-on galaxies, i.e., with inclination larger than 70$^o$, to avoid possible issues related with highly inclined galaxies. All data were analyzed using the Pipe3D pipeline \citep{sanchez+2016a,lacerda22}, a tool aimed to separate the stellar population and emission line properties of the galaxies. Using this tool we retrieve for each galaxy: (i) the integrated stellar mass, M$_*$, derived from via the decomposition of the stellar population spectra in a set of single-stellar populations spectra, which allows to estimate the average mass-to-light ratio; (ii) the integrated the star-formation rate, estimated using the dust-corrected H$\alpha$ luminosity, adopting the calibrator proposed by \citet{kennicutt1998}; and finally (iii) an estimation of the molecular gas mass (M$_{gas}$) based on the dust attenuation, using a modification of the calibrator proposed by \citet{jkbb20}. All those quantities were derived using the procedures described in detail in \citet{sanchez21} and \citet{sanchez22}. In addition we retrieve the [OIII]/H$\beta$, [NII]/H$\alpha$ line ratios and the EW(H$\alpha$) measured at the effective radius. Using those data we define a sample of star-forming galaxies (SFGs) following \citet{ARAA}, i.e., galaxies that are located below the \citet{kewley01} demarcation line in the BPT \citep{baldwin81} diagram with an EW(H$\alpha$)$>$6\AA.  Figure \ref{fig:sample} shows the distribution along the SFR-M$_*$ plane of the 485 (non highly inclined) galaxies selected from the AMUSING++ compilation, together with the SFGs sub-sample, that comprises 297 galaxies. It is clearly seen that this compilation comprises galaxies of any star-formation stage, covering a wide range of stellar masses, and presumely a wide range of galaxy morphologies \citep[as shown in][]{carlos20}. Finally, it is clear that our selected sub-sample of SFGs follow the expected trend along this diagram, i.e., the SFMS.

\begin{figure}
    \centering
    \includegraphics[width=\textwidth]{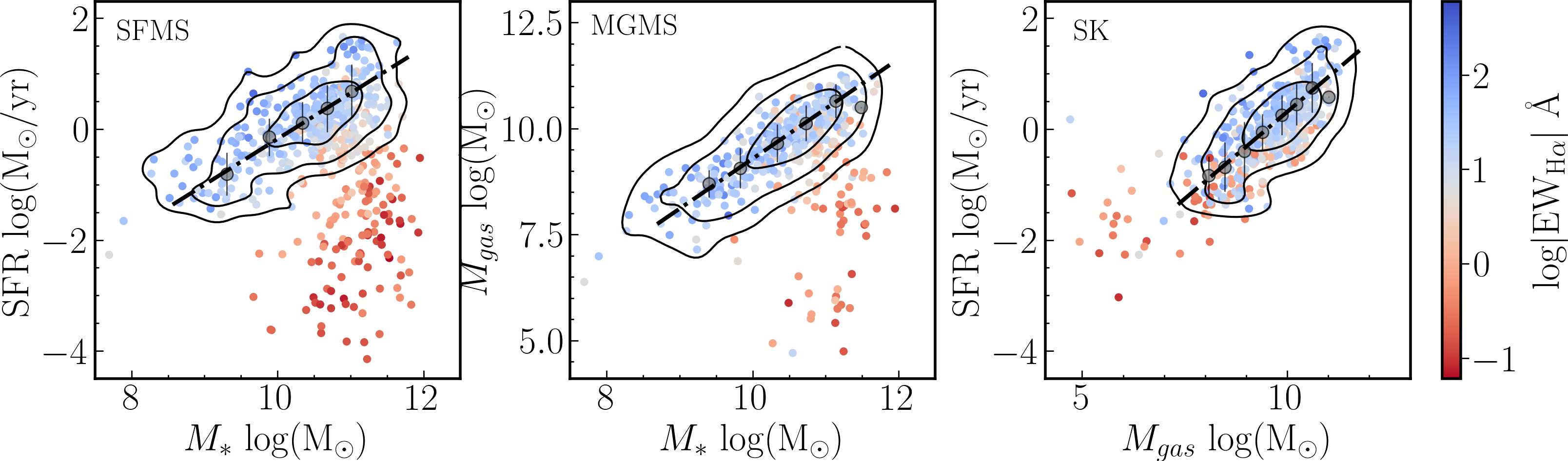}
    \caption{Distribution of the global extensive parameters: (i) SFR as a function of M$_*$ (left panel), (ii) M$_{gas}$ as a function of M$_*$ (central panel) and (ii) SFR as a function of M$_{gas}$ (right panel). In each panel each  solid-circle corresponds to one AMUSING++ galaxy, color-coded by the EW(H$\alpha$) at the effective radius. The contours trace the density distribution of the SFGs (encircling 95\%, 80\% and 40\% of this subsample, respectively). Grey solid-circles corresponds to the average values in bins of 0.2 dex in the x-axis, with the errorbars tracing the 1$\sigma$ distribution. Finally the dashed-lines correspond the best fitted relations: SFMS, MGMS and SK, in each case.} 
    \label{fig:ext}
\end{figure}

\section{Analysis and Results}

Following \citet{mariana19} we first explore the global extensive relations between the three retrieved parameters (SFR, M$_*$ and M$_{gas}$), analyzing the distribution and relation between the SFR-M$_*$ (SFMS), M$_{gas}$-M$_*$ (MGMS) and SFR-M$_{gas}$ (SK). Figure \ref{fig:ext} show such distributions for the full sample of galaxies, highlighting the location of SFGs. As expected, in the three cases it is observed clear linear trends that define the well known relations. To parametrize them we perform a binning of the data in the x-axis, deriving the average and standard deviation in bins of 0.2 dex. The location of the binned data is shown in the different panels too. Then, we perform a simple linear regression for the binned data to derive the best fitted relation between both pairs of parameters. Equations 3.1, 3.2, and 3.3 show the results of this analysis. Errors were derived by performing a
simple Monte-Carlo iteration on the original data, propagating the errors of each derived parameter.


\begin{equation}\label{eq:1a}
    log(SFR)=(0.83\pm 0.22)log(M_*)-(8.43\mp 0.93)
\end{equation}
\begin{equation}\label{eq:1b}
    log(M_*)=(1.18\pm 0.21)log(M_{gas})-(2.52\mp 0.83)
\end{equation}
\begin{equation}\label{eq:1c}
    log(SFR)=(0.63\pm 0.19)log(M_{gas})-(5.97\mp 0.67)
\end{equation}

As expected we found three clear tight correlations between the analyzed extensive parameters. In all cases the slopes are near one, or slightly lower, by compatible within $\sim$2$\sigma$ with this value. The largest deviation from this value is reported for the SK-law, most probably due to the more limited dynamical range of the two parameters involved in this relation.

\begin{figure}
    \centering
    \includegraphics[width=\textwidth]{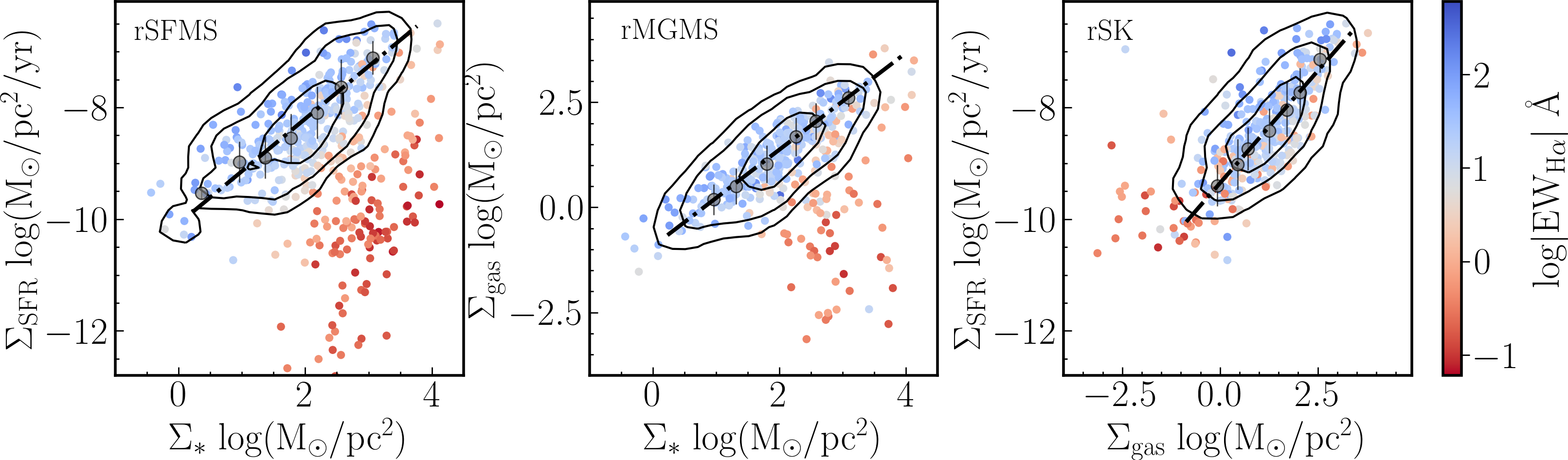}
    \caption{Distribution of the global intensive parameters: (i) $\Sigma_{\rm SFR}$ as a function of $\Sigma_\star$ (left panel), (ii) $\Sigma_{\rm gas}$ as a function of $\Sigma_\star$ (central panel) and (ii) $\Sigma_{\rm SFR}$ as a function of $\Sigma_{\rm gas}$ (right panel). In each panel each  solid-circle corresponds to one AMUSING++ galaxy, color-coded by the EW(H$\alpha$) at the effective radius. The contours trace the density distribution of the SFGs (encircling 95\%, 80\% and 40\% of this subsample, respectively). Grey solid-circles corresponds to the average values in bins of 0.2 dex in the x-axis, with the errorbars tracing the 1$\sigma$ distribution. Finally the dashed-lines correspond the best fitted global intensive relations relations: rSFMS, rMGMS and rSK, in each case.} 
    \label{fig:int}
\end{figure}

Once derived the extensive relations, we repeat the analysis using the global intensive relations (\Ssfr, \Sst and \Sgas). To derive those parameters we follow \citet{sanchez21}, dividing the corresponding extensive parameter by the effective area covered by each galaxy (defined as 4$\pi$R$^2_e$, i.e., the area encircled in 2 effective radius). Figure \ref{fig:int} show the distribution of parameters and the results of the binning and fitting analysis, with the best fitted linear relations listed in Equations 3.4, 3.5, and 3.6.


\begin{equation}
    log({\Sigma}_{SFR})=(0.94\pm 0.22)log({\Sigma}_{*})-(10.08\mp 0.35)
\end{equation}
\begin{equation}
    log({\Sigma}_{gas})=(1.15\pm 0.19)log({\Sigma}_{*})-(0.92\mp 0.30)
\end{equation}
\begin{equation}
    log({\Sigma}_{SFR})=(0.80\pm 0.18)log({\Sigma}_{gas})-(9.34\mp 0.22)
\end{equation}

As expected the three distributions show a clear tight relation between the involved pair of parameters, correlations that a tighter than the extensive ones, with a lower error for both the slopes and zero-points. Furthermore, the slopes are in all cases more near to one, within less than 1$\sigma$ in the case of the rSFMS and rMGMS relations and just at 1$\sigma$ for the rSK one.

\begin{table}
\small
\centering
\caption{Results of the analysis of global intensive relations rSFMS, rSK y rMGMS}
\label{tab:2D}
\begin{center}
\begin{tabular}{lrrrrrrr}
\hline
  \multicolumn{1}{c}{References} & \multicolumn{1}{c}{$\beta$} & \multicolumn{1}{c}{$\alpha$} &  \multicolumn{1}{c}{r$_c$} & \multicolumn{1}{c}{$\sigma_{\rm obs}$} & \multicolumn{1}{c}{$\sigma_{exp}$} &
  \multicolumn{1}{c}{\#} &  \multicolumn{1}{c}{\#}\\
  \multicolumn{6}{c}{} &
  \multicolumn{1}{c}{Gal.} &
  \multicolumn{1}{c}{SFG}\\
  \hline
    \multicolumn{8}{c}{rSFMS} \\
  \hline
  AMUSING++   & -10.08$\pm$0.35 & 0.94$\pm$0.22 & 0.79 & 0.466 & 0.290 & 485 & 297\\
  \hline
  {\it Sa21a(EDGE)}   & -10.10$\pm$0.22 & 1.02$\pm$0.16 & 0.68 & 0.266 & 0.190 & 126 & 12667\\
  Sa21a(CALIFA) & -10.27$\pm$0.22 & 1.01$\pm$0.15 & 0.85 & 0.244 & 0.192 & 941 & 533\\
  Sa21a(APEX)   &  -9.78$\pm$0.30 & 0.74$\pm$0.21 & 0.76 & 0.226 & 0.211 & 512 & 251\\
    \hline
  {\it Wu13}      & -8.4$^1$ & 0.95 &  & \\
 {\it CD16}    & -10.19$\pm$0.33 & 0.72$\pm$0.04 & 0.63 & 0.16 \\
 {\it Li19}        & -10.54$\pm$0.11 & 1.19$\pm$0.01 &    & 0.25 &   & 14 & 5383$^{*}$
  \\
 {\it CD19}    & -10.48$\pm$0.69 & 0.94$\pm$0.08 & 0.62 & 0.27 &  & 2737 & $\sim$500K\\
 Sa21    &-10.35$\pm$0.03 & 0.98$\pm$0.02 & 0.96 & 0.17 &   &1512 & $\sim$3M\\
{\it El20}   &-10.07$\pm$1.44 & 1.03$\pm$0.17 & 0.57 & 0.28-0.39 &   & 28 & $\sim$15035$^{*}$\\      
\hline
    \multicolumn{8}{c}{rSK} \\
  \hline
  AMUSING++   & -9.34$\pm$0.22 & 0.80$\pm$0.18 & 0.8 & 0.478 & 0.294 & 485 & 297\\
  \hline
  {\it Sa21a(EDGE)}   & -9.01$\pm$0.14 & 0.98$\pm$0.14 & 0.73 & 0.249 &  0.216 & 126 & 12667\\
  Sa21a(CALIFA) & -9.01$\pm$0.16 & 0.95$\pm$0.21 & 0.77 & 0.293 &  0.297 & 941 & 533\\
  Sa21a(APEX)   & -8.84$\pm$0.24 & 0.76$\pm$0.27 & 0.70 & 0.294 &  0.228 & 512 & 351\\
    \hline
  {\it Bo17}    & -9.22            & 1.00          &      &      & & 104 & $\sim$5000\\
  {\it Li19   }     & -9.03$\pm$0.06   & 1.05$\pm$0.01 &      & 0.19 & & 14 & 5383$^{*}$\\
{\it  El20}   & -8.87$\pm$0.66 & 1.05$\pm$0.19 & 0.74  & 0.22-0.32 &   & 28 & $\sim$15035$^{*}$\\  
\hline  
    \multicolumn{8}{c}{rMGMS} \\
  \hline
 AMUSING++   & -0.92$\pm$0.30 & 1.15$\pm$0.19 & 0.85 & 0.528 & 0.283 & 485 & 297\\
 \hline
 {\it  Sa21a(EDGE)}  & -0.91$\pm$0.16 & 0.93$\pm$0.11 & 0.68 & 0.218 & 0.209 & 126 & 12667\\
 Sa21a(CALIFA) & -1.12$\pm$0.27 & 0.93$\pm$0.18 & 0.74 & 0.276 & 0.288 & 941 & 533\\
  Sa21a(APEX)  & -0.70$\pm$0.37 & 0.73$\pm$0.24 & 0.73 & 0.234 & 0.212 & 512 & 251\\
    \hline
 {\it Li19}        & -0.59$\pm$0.08  & 1.10$\pm$0.01 &    & 0.20 & & 14 & 5383$^{*}$\\
 {\it BB20}    & -0.95           & 0.93          &    & 0.20 & & 93 & $\sim$5000\\
 {\it El20}   & -0.99$\pm$0.13  & 0.88$\pm$0.15 & 0.72 & 0.21-0.28 &   & 28 & $\sim$15035$^{*}$\\
\hline
\end{tabular}
\caption{Zero-point ($\beta$) and slope ($\alpha$) of the three intensive global relations explored along this study (rSFMS, rSK and rMGMS), together with the values reported for a selection of similar relations extracted from the literature: \cite{sanchez2021b}(Sa21a), \cite{wuyts13}(Wu13), \cite{mariana16}(CD16), \cite{lin19}(Li19), \cite{mariana19}(CD19), \cite{sanchez21}(Sa21) y \cite{ellison21a}(El20), \cite{sanchez21}(Sa21), \cite{bolatto17}(Bo17), \cite{lin19}(Li19) and  \cite{jkbb20}(BB20) y \cite{ellison21a}(El20). In addition
we include the correlation coefficient ($r_c$), the standard deviation of
the distribution of data before ($\sigma_{\rm obs}$) and after applying the best fitted linear relation ($\sigma_{\rm exp}$), together with the number of galaxies or number of individual spatial elements considered in the analysis. In italics we highlight the results from spatially resolved data, and in non-italics are indicated the explorations based on global intensive parameters.}
\end{center}
\end{table}

\section{Conclusions}

Table \ref{tab:2D} present a summary of the results together with similar values
reported in the literature for both global intensive relations (Fig. \ref{fig:int}),
and local/resolved ones. From this comparison it is seen that:

\begin{itemize}

\item Global intensive parameters present a niche of opportunity to explore the physics of star-formation in a more coherent way than extensive ones, providing tighter and better defined relations (as the galaxy extension and aperture effects are limited).

\item Global intensive relations can be directly compared with local/resolved ones (at least down to 1 kpc), showing similar trends both qualitatively and quantitatively, suggesting that they are originated by the same physical processes.

\item There are hints of the evolution of the local/resolved relations that are evident when comparing results from sample at different redshits (e.g. \citet{wuyts13} vs. the rest of the data)

\item If there is any secondary relation or a hierarchy among the relations, it is required to explore them on the tighter and best defined set of data.

\end{itemize}

In this regards, we need to highlight that it is still an open question
which is of the explored relations is more fundamental than the previous one (if any).
The nature of the scatter described in the three relations
and the existence of possible secondary relations with other
parameters, that may drive this scatter, is indeed and  important topic of study. 
As indicated before, it could be the case that these relations depends on the 
morphology, that may affect the star-formation efficiency (SFE).
Some authors have reported that the rSFMS preent a secondary relation that depends
on the SFE \citep[][using the ALMAQUEST data]{ellison20} or the gas fraction \citep{colombo20}. This has open the seek for a more fundamental relation that
involves pairs of parameters to trace the SFR \citep[e.g.][]{shi18,dey19,jkbb21a}.
However, other results indicate that most probably those secondary relations
are pure mathematical artifact introduced by correlation between errors \citep{sanchez2021b}. The use of a larger samples, the tighter and better defined relations (i.e., the intensive ones), and better methods to derive the parameters
(in particular for the molecular gas), seem to be required approach to perform future explorations.

\acknowledgements

SFS thanks the support by the PAPIIT-DGAPA IG100622 project.
Based on data obtained from the ESO Science Archive Facility.


\bibliographystyle{aasjournal}
\bibliography{my_bib}

\end{document}